 \documentclass{article}
 \usepackage{anysize}
 \usepackage[latin1]{inputenc}
 \usepackage{graphicx}
 \usepackage{pdflscape}
 \usepackage{float}
 \usepackage{amsmath, amsthm, amssymb}
 \usepackage[T1]{fontenc}
 \usepackage{authblk}

 \title{A Framework for Automated Cell Tracking in Phase Contrast Microscopic Videos based on Normal Velocities}
 \author{Michael M\"{o}ller}
 \author{Martin Burger\footnote{Corresponding author. Email address: martin.burger@uni-muenster.de, phone: +49 251 83 33793, fax: +49 251 83 32714}}
 \affil{Institut für Numerische und Angewandte Mathematik, Westfälische Wilhelms-Universität, Einsteinstrasse 62, 48149 Münster}
 \author{Peter Dieterich}
 \affil{Medizinische Fakultät Carl Gustav Carus, Technische Universität Dresden, Fetscherstrasse 74, 01307 Dresden}
 \author{Albrecht Schwab}
 \affil{Institut für Physiologie II, Westfälische Wilhelms-Universität, Robert-Koch-Strasse 27b, 48149 Münster}
 
\begin{document}
\maketitle
\vspace{3cm}
\hrule
\vspace{1cm}
\textbf{Abstract.}
This paper introduces a novel framework for the automated tracking of cells, with a particular focus on the challenging situation of phase contrast microscopic videos. Our framework is based on a topology preserving variational segmentation approach applied to normal velocity components obtained from optical flow computations, which appears to yield robust tracking and automated extraction of cell trajectories. In order to obtain improved trackings of local shape features we discuss an additional correction step based on active contours and the image Laplacian which we optimize for an example class of transformed renal epithelial (MDCK-F) cells. We also test the framework for human melanoma cells and murine neutrophil granulocytes that were seeded on different types of extracellular matrices. The results are validated with manual tracking results. 

\vspace{0.5cm}
\textit{Key words.} Cell Tracking, Phase Contrast Microscopy, Optical Flow, Active Contours, Melanoma Cells
\vspace{1cm}
\hrule

%



\section{Introduction}
\label{intro}
Nearly all cells in the human body are moving at some point during their lifetime thereby contributing to the maintenance of the integrity of the human body and to medically important pathologies. Two prominent examples for the medical relevance of cell migration are the formation of tumor metastases and the immune response. Tumor cell migration is a critical step of the metastatic cascade that leads to the spread of tumor cells within the body (\cite{FriedlWolf03, GuptaMassague06}). Thus, tumor cell migration is an important factor in determining the prognosis of cancer patients since patient outcome largely depends on the presence or absence of tumor metastases. Similarly, the ability of immune cells to migrate towards sites of infection is indispensable for clearing invading pathogens from the organism. On the other hand migration of immune cells towards foci of inflammation may contribute to the clinical symptoms of e.g. an autoimmune disease (\cite{WongHeitKubes10}).

	On a cellular level, cell migration represents the integration of a network of many different molecular mechanisms within the cell. Such mechanisms include among others the dynamic remodeling of the cytoskeleton (\cite{OlsonSahai09}), coordinated formation and release of cell-matrix contacts (\cite{RoseAlonGinsberg07}), and the activity of ion transporters and channels (\cite{Schwabetal07}). These molecular mechanisms are regulated by networks of intracellular and extracellular signaling molecules so that the migratory activity can be adapted to the respective requirements and migrating cells can respond to external cues. The contribution of individual components of the cellular migration machinery can be assessed by quantitatively evaluating the migratory behavior of individual cells, i. e. by determining speed, translocation and shape of single cells. 

	The "bottleneck" of such a detailed analysis is the extremely time-consuming segmentation of cells in image stacks of time-lapse videomicroscopic migration experiments. Thus, there is an urgent need for the development of automated sophisticated cell tracking software. Such a system should be able to detect not only the movement of the cell center but also to quantitatively assess changes of cell morphology during the process of cell migration. Large data sets obtained from automated cell tracking covering several time scales (from seconds to hours) could also be used for system biological modeling of cell migration. Finally, the automation of the image processing can facilitate the application of high-throughput screening regimes to single cell videomicroscopic migration experiments.

We propose a semi-automatic tracking method with minimal user interaction. The user only has to determine the rough positions of cells in the first frame. The method is based on a two step algorithm including a rough tracking using motion information and a contour refinement based on the gray level image. In this paper we will first summarize existing cell tracking techniques. We briefly describe the challenges in cell tracking (Section \ref{datachallenge}) as well as the data acquisition process for our cell migration experiments (Section \ref{experiments}) before we discuss how to tackle the problems and propose a tracking method in Section \ref{ourmethod}. Numerical results and an evaluation of our method in comparison to manual tracking will be given in Section \ref{results} and we will draw conclusions in Section \ref{conclusion}.	

\section{State of the Art}
\label{stateofart}

There are many ways and methods for tracking movement.  For a general summary of object tracking methods we refer to \cite{generaltracking}. A good summary of the different techniques specifically developed for cell tracking can be found in \cite{MiruraBook}. Although `cell tracking' is a fixed technical term there is no general `cell tracking' method, since the cell images can be very different depending on the cell type, the environment, and the image capture modality the experiment is recorded with. Furthermore, there are many relevant papers in the computer vision comunity addressing video tracking in general.  However, we will try to summarize some approaches that usually have to be adopted to the data they are used for. We will structure the summary based on the techniques used for the tracking. 

The simplest way besides manual segmentation is thresholding the data, which is not applicable in most cases since a big intensity difference between the objects of interest and the background as well as a noise free image are needed. Thresholding might be applicable to tracking of
fluorescently labeled particles on a black background.
 
Many methods use templates to perform pattern matching. A template is moved over the image domain and compares the patch at the current position with the template e.g. via cross-correlation (\cite{CCmatching}). Other methods try to fit Gaussian curves to their signal and use the peak position as the position of the object(\cite{gaussref2, gaussref1}). Having segmented the cells in each frame the trajectory is recovered by searching for the nearest centroid position in the previous frame \cite{squareerrordist} or by using graph based methods (\cite{graphbasedconnection, MiruraBook}). The approach of Debeir et al. (\cite{DebeirMeanShift}) was to use a multikernel mean shift algorithm based on the assumption that the cell interior and the exterior close to the cell boundary have different gray levels. 

Another class of methods are motion estimation methods. They try to calculate the so-called optical flow and segment the video based on the object's movement. The most common methods are Horn-Schunck (\cite{hornschunck}) and Lucas-Kanade (\cite{lucaskanade}). A comparison of general optical flow algorithms can be found in \cite{BarronOptFlowPerform}. An applications of optical flow to cell tracking is presented in \cite{optflowtrack1}. The optical flow equation can also be used for registration, which is done in \cite{Hand09} for tracking many cells in a phase-contrast video microscopy sequence. The tracking yields good results but only for about 50 frames. After that an accumulation of error causes to large deviations from the actual cells centroid position.

Recently, active contour models or 'snakes' attracted a lot of attention for general image segmentation. This approach minimizes an energy depending on the segmentation curve, where a low energy corresponds to a curve with the desired properties. Typically, these methods are driven by the data in some feature space and make a regularity assumption on the smoothness of the curve. They can also include a-priori information about the shape, volume or position of the object of interest. Our method uses the latter approach with two different energies in two different steps. The most prominent representatives of these kind of models are Mumford-Shah, Chan-Vese, and Geodesic Active Contour models or Snakes. A very good summary of active contour cell tracking methods can be found in \cite{ZimmerSummary}. There are many examples for active contour methods in cell tracking. Ray et al. track leukocytes with active contours under size and shape constraints \cite{RayTrackingLeukocytes}. In \cite{WangTextureSnakes} a statistical model is used to train the algorithm estimating the typical texture of cell boundaries and in \cite{ZimmerParamActCont} a local average filter is used to detect low contrast boundaries. Recent research on using level set functions for cell tracking in fluorescent microscopy can also be found in the work group of Meijering (see \cite{Dzyubachyk10, Dzyubachyk08} and the references therein).

Other hybrid methods using active contours and motion information have been proposed. \cite{ParagiosACTracking} designed a method where the segmentation is driven by a statistical motion estimation force as well as a spatial gradient dependent force. Bascle et al. \cite{Bascle94} present a tracking method using snakes with an additional motion constraint, whereas the motion is estimated in a multi scale approach using Gaussian pyramids. In \cite{Papin00} motion estimates are used to design parts of the force driving a level set function.

Due to the big differences in different cell images many methods have to be developed data specific and therefore address different problems. A performance comparison of single particle tracking methods in fluorescent microscopy can be found in \cite{CheezumComparison}. A difficult problem in tracking multiple cells is the danger of wrongly merging adjacent cells, which was tackled in \cite{ZhangMultiLS, NathMultiphaseLS, ZimmerSummary} by using multiple level set functions with a repulsion term. As we will describe in more detail in Section \ref{datachallenge} the cells in phase contrast microscopy are often surrounded by bright spots, so-called halos, which are artifacts from the image acquisition process and make an accurate determination of the cell boundary difficult. For the case of a given rough segmentation, Ersoy et al. \cite{ErsoyNormalDirectionRefinement} addressed this problem and proposed a segmentation refinement based on geodesic active contours. The initial contour is assumed to be outside the cell of interest. It is then moved inwards until the gray value in normal direction changes from bright to dark. The second processing step of our approach will be closely related to this method.  

\section{Challenges in Phase Contrast Microscopy Videos}
\label{datachallenge}
\begin{figure}[ht]
			\centering
			\includegraphics[width=14cm]{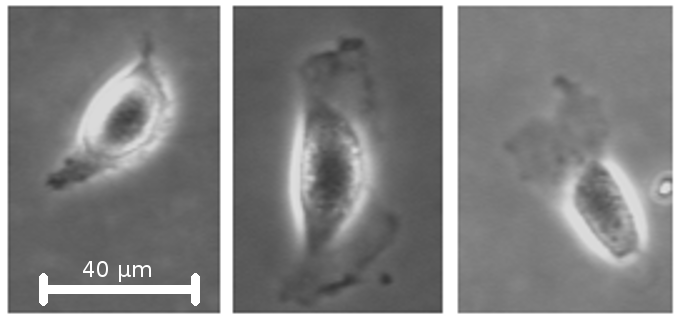}
			\caption{Phase contrast micrographs of MDCK-F cells from time-lapse videomicroscopic migration experiments. Example of the halos of three cells.}
			\label{halos}
\end{figure}

Phase contrast microscopy, which is applied in this study offers the
advantages that - as opposed to fluorescence microscopy - it is essentially
devoid of any phototoxicity and that cells can be observed in their native
state without the necessity of dye-loading or transfection with fluorescent
proteins. It thereby allows the monitoring of cell migration at high time
resolution for extended time periods without any side effects on the cellular
behavior. However, the downside of phase contrast microscopy is that it poses 
specific challenges to the tracking algorithms. Due to the specific illumination technique gray levels within a given cell or in its immediate vicinity may greatly vary during the course of an experiment. Thus, the gray level of the cell can be higher or lower than that of the background. Moreover, some parts of the cells will have gray levels that are almost indistinguishable from that of the background. Yet another difficulty in determining the cell contours in phase contrast images is that the rear part of migrating cells is frequently surrounded by a bright "halo". Figure \ref{halos} shows a close up on three cells with different halos. As we can see the halo does not follow any regular model. It can be inside as well as outside the actual cell boundary. For these images even a human analyst can only guess the actual contour of each cell. This is a well known problem and causes different manual segmentations on the exact same dataset to yield relatively different results as we will show in the results section. Neither omitting all white spots nor including them nor looking for certain intensity changes in the halo yields the true contour. 

\begin{figure}[ht]
			\centering
			\includegraphics[width=15cm]{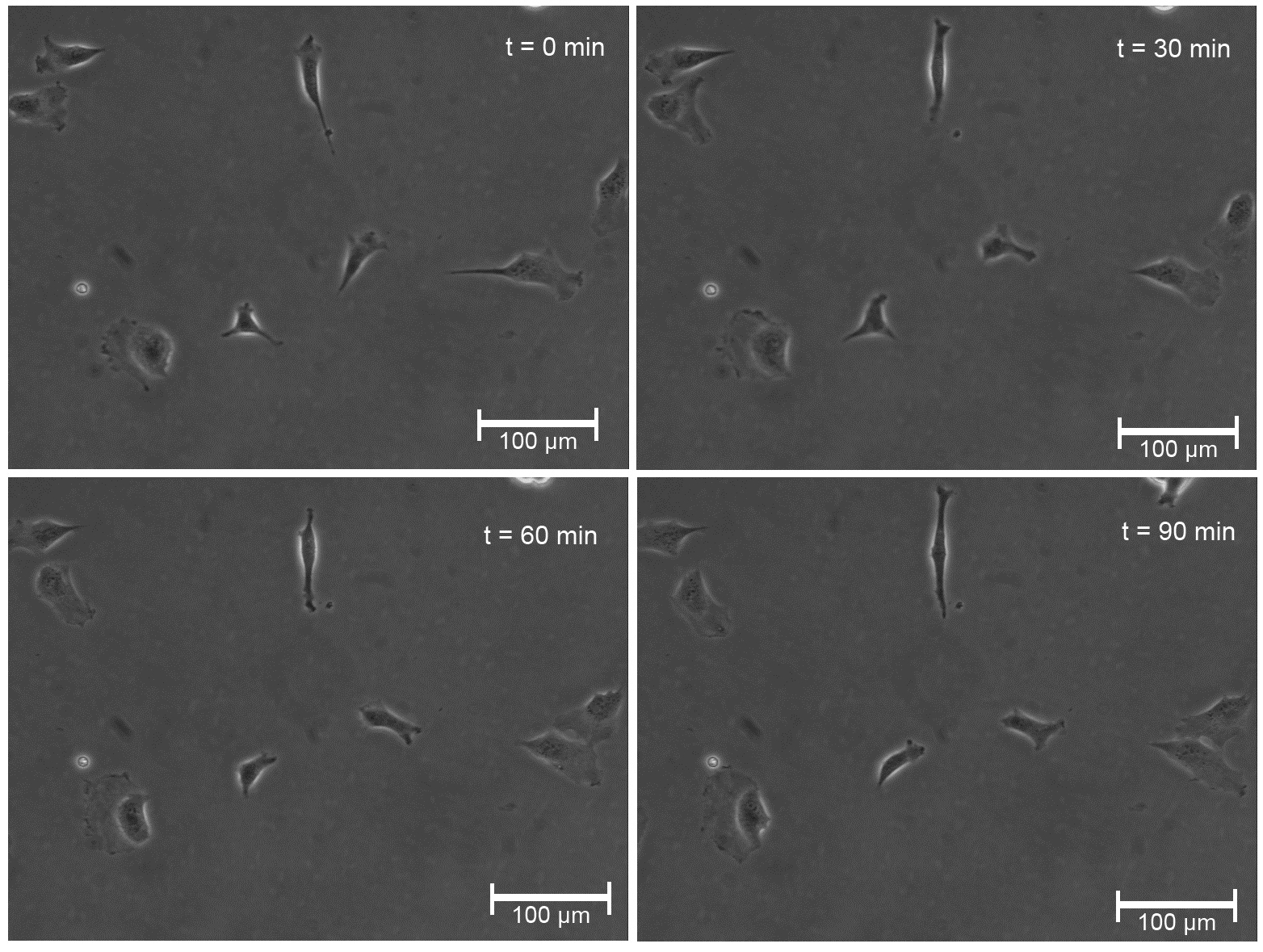}
			\caption{Phase contrast micrographs of MDCK-F cells from time-lapse videomicroscopic migration experiments. The frames of this experiment will be used to illustrate our cell tracking technique in Figures \ref{normalvel}, \ref{wrongtopologysegementation}, \ref{initcontdependence}, \ref{topocomparison} and \ref{gmap}.}
			\label{data}
\end{figure}

Figure \ref{data} shows four frames of the data we are working with and illustrates a further complication. The cells have very different shapes and undergo strong deformations over time. Finally, the background of the images is not always as homogeneous as in Fig. \ref{data}. It may also contain dirt, dead cells, or structures derived from the substrate (extracellular matrix) the cells are moving on. These structures may also move as a consequence of the mechanical forces exerted by migrating cells. All these structures of the background should not be part of the segmentation.

\section{Cell migration experiments and data acquisition}
\label{experiments}
In the following we briefly summarize the type of experiments the videomicroscopic data was acquired with. Cell migration was performed as described previously (\cite{Schwabetal05, Kraehlingetal09}). Cells were plated in culture flasks and placed in heated chambers ($37°C$) on stages of inverted phase contrast microscopes (Axiovert 40c and 20, Zeiss). Depending on the cell type cell migration was recorded in 5 s (murine neutrophil granulocytes), 1 min (MDCK-F cells), or 10 min intervals (MV3 melanoma cells) for up to 5 h using video cameras (Models XC-ST70CE and XC-77CE, Hamamatsu/Sony, Japan) and PC vision frame grabber boards (Hamamatsu, Herrsching, Germany). Acquisition of images was controlled by the HiPic and WASABI software (Hamamatsu). For manual segmentation of the cell contours and analysis of cell migration AMIRA software (Visage Imaging, Berlin, Germany) was applied in conjunction with JAVA and ImageJ programs developed by ourselves. The cell contours served as the basis for further analysis (\cite{Dieterichetal08}). The x- and y- coordinates of the cell center ($\mu m$) were determined as geometric mean of equally weighted pixel positions within the cell outlines. Cell areas ($A$; $ \mu m^2$) are defined as the appropriately scaled sum of all pixels inside the cell´s contours. The structural index (SI), a shape factor, was calculated as follows: $SI = (4 \pi A)/p^2$, where $p$ denotes the cell perimeter.

\section{The Proposed Cell Tracking Framework}
\label{ourmethod}
We propose a semi-automatic tracking algorithm with minimal user interaction based on two general steps, a rough segmentation using motion information and a contour refinement based on local gray value variations. We transfer the technique of topology preserving level set methods as proposed in \cite{topologypreservation} for geodesic active contours of a single level set function to segmenting motion based images with a Chan-Vese type of energy with multiple level set functions and an additional volume constraint. Although we developed the algorithm for transformed rual epithelial (MDCK-F) cells we will show that the approach of tracking cells by their normal velocity works for a much wider class of microscopic cell images. 

The contributions in this paper include the following: We present a complete cell tracking framework for obtaining not only the cells centroid but a contour of each cell enabling the calculation of area and shape of the cell. The centroid position in particular can be determined in very good accuracy without accumulation of error. According to all our experiments cells are never lost and the tracking results do not degrade over time.  To the best knowledge of the authors, the volume constraint as well as the application of topology preservation techniques to the Chan-Vese energy are novel. Furthermore, the evaluation technique of comparing the precision of our method to the average deviation between different human observers on the same data set has not yet been exploited.

The proposed segmentation framework can be divided into the following parts:
\begin{enumerate}
	\item Calculating the normal velocity
	\item Segmenting the normal velocity image with a Chan-Vese method including a volume constraint
	\item  Preserving the topology throughout the tracking procedure
	\item Refining the contour with a geodesic active contour model
\end{enumerate}
In the following subsections we will describe each part in detail.

\subsection{Calculating the normal velocity}
The origin of our first tracking step lies in the theory of optical flow calculation. The idea behind optical flow is to estimate the velocity of objects in a video scene by looking at two (or more) successive frames. Let $f(x,t)$ be a continuous representation of a video with the two dimensional space variable $x=(x_1, x_2) \in \Omega \subset \mathbb{R}^2$ and the time variable $t \in \mathbb{R}$. We are given two frames, $f(x,t)$ and $f(x,t+ \Delta t)$ and search for the velocity that objects in $f(x,t)$ were moved with to obtain $f(x,t+\Delta t)$. To derive an equation for the desired velocity we make the standard assumption that the intensity values of our objects stay the same during the motion, i.e. 
\begin{eqnarray}
\label{grayvalconst}
f(x,t) = f(x+ \Delta x, t+\Delta t).
\end{eqnarray}
 We use a first order Taylor approximation for the latter term to obtain
\begin{eqnarray}
f(x+ \Delta x, t+\Delta t) \approx f(x,t) + \Delta t \frac{\partial}{\partial t} f(x,t) + \nabla f(x,t) \cdot \Delta x
\end{eqnarray}
or, with $\vec{v} = \frac{\Delta x}{\Delta t}$,
\begin{eqnarray}
f(x+\Delta x, t+\Delta t) \approx f(x,t) +  \Delta t \frac{\partial}{\partial t} f(x,t) + \Delta t \; \nabla f(x,t) \cdot \vec{v}.
\end{eqnarray}
Using the gray value constancy assumption (\ref{grayvalconst}) we get
\begin{eqnarray}
0 \approx  \frac{\partial}{\partial t} f(x,t) + \nabla f(x,t) \cdot \vec{v}.
\end{eqnarray}
Typically, there are two spatial dimensions such that $\vec{v}$ is a vector in $\mathbb{R}^2$. The above system therefore is underdetermined and the calculation of the velocity is ill-posed. Hence, one needs to regularize (e.g. like in the Horn-Schunck method) or assume constant velocity for small patches of the image (e.g. like in the Lucas-Cascade method). However, calculating the absolute value of the normal velocity is well defined for $|\nabla f(x,t)| \neq 0$. In this case we can divide by this quantity and denote the normal velocity by $v_n$. This leads us to
\begin{eqnarray}
\label{normalvelformu}
| v_n | \approx \frac{|\frac{\partial}{\partial t} f(x,t)|}{|\nabla f(x,t)|_{\epsilon}}  ,
\end{eqnarray}
where we use $|\nabla f(x,t)|_{\epsilon} = \sqrt{ (\partial_{x_1} f)^2 +(\partial_{x_2} f)^2+ \epsilon^2}$ for a small $\epsilon$ to extend the normal velocity for zero gradients. For our numerical experiments we discretized equation (\ref{normalvelformu}) using central differences for the time derivative 
\begin{eqnarray}
\frac{\partial}{\partial t} f(x = (i \Delta x_1, j \Delta x_2), t=k \Delta t) \ \approx \  \frac{f_{i,j,k-1} - f_{i,j,k+1}}{2 \Delta t},
\end{eqnarray}
without prior preprocessing or frame alignment, where $i,j$ denote the two discrete spatial coordinates and $k$ the current frame number. We used $\Delta x_1 = \Delta x_2 = \Delta t =1 $ for our video. For the spatial derivative we used a backwards Euler discretization.

Tracking based on normal velocities has been proposed before by Irani, Rousso and Peleg in \cite{Irani94}. They propose to calculate the normal velocity at a pixel by averaging over a small neighborhood and classify pixels into moving or non-moving in a hierarchical approach based on normal velocities and the condition number of the optical flow matrix on different levels of a Gaussian pyramid. 

\begin{figure}[ht]
			\centering
			\includegraphics[width=15cm]{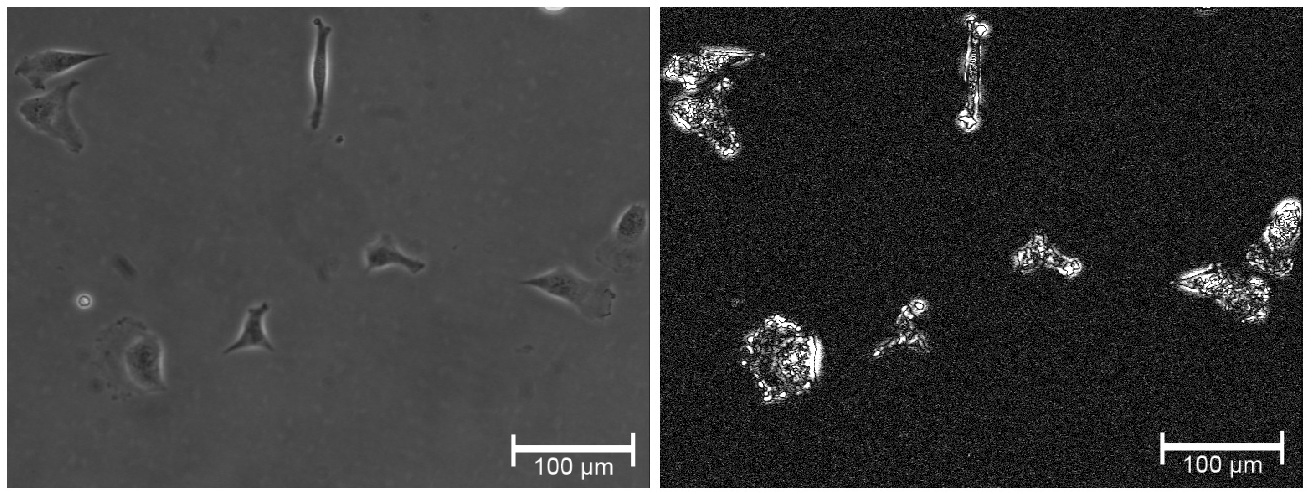}
			\caption{Absolute value of the normal velocity in a cell frame according to Formula (\ref{normalvelformu}).}
			\label{normalvel}
\end{figure}

The averaging over a pixels neighborhood as well as lower levels of a Gaussian pyramid help to deal with noise. However, in our case the contrast is low and the motion rather small, so that an averaging did not lead to an improvement of the results. Figure \ref{normalvel} shows an example of a normal velocity image. We can see that although we have quite a lot of noise the cells can clearly be identified as the moving object, i.e. parts of the image with high normal velocity. 

\subsection{Segmenting the normal velocity image with a Chan-Vese method}
We need a method to segment the bright normal velocity parts despite the noise and the possible small darker areas in the cell interior in a `meaningful' way. Simple thresholding would lead to thousands of unconnected components. We instead use the variational segmentation method of Chan and Vese, (\cite{ChanVese}), to obtain larger connected components. The general idea of variational segmentation methods is to design an energy model, where a low energy value corresponds to a good segmentation. The final contour is then obtained by finding the argument that minimizes the energy. 

In our approach we represent the segmentation by so-called level set functions. The sign of each level set function at each pixel determines its affiliation to each cell, meaning for $N$ cells we would like to obtain functions $\phi_j, \ j=1,...,N$ such that
\begin{eqnarray}
\phi_j(x)&<&0, \ \ \text{if x is inside cell $j$}, \nonumber \\
\phi_j(x)&>&0, \ \ \text{if x is outside cell $j$}, \nonumber \\
\phi_j(x)&=&0, \ \ \text{if x is on the boundary of cell $j$}. \nonumber
\end{eqnarray}
We can address the inside and outside of the cells by applying the Heaviside function
\begin{eqnarray}
H(\phi) = 0  \ \ \text{for } \phi \leq 0, \nonumber \\
H(\phi) = 1  \ \ \text{for } \phi > 0,
\end{eqnarray}
which means that $H(\phi)$ is the indicator function of the cells exterior and $1-H(\phi)$ the one of the cells interior. The background, i.e. the intersection of all cell exteriors, can be expressed as $ (\prod_{j=1}^{N} H(\phi_j))$.

We have seen in the previous subsection that the cells can be identified with image regions of high normal velocity. The outside (background of the image) should have low normal velocity. Therefore, it is reasonable to look for image partitions such that each cell interior has a typical high normal velocity, $c_{inside}^j$, and their common background has a low normal velocity, $c_{outside}$. Denoting the image domain by $\Omega$ and the absolute value of the normal velocity by $|v_n|$, we wish to minimize
\begin{eqnarray}
E_{data}(\{\phi_j\}_{j=1,...N}, \{c_{inside}^j\}_{j=1,...N}, c_{outside})& =& \lambda_1 \sum_{j=1}^{N}  \int_{\Omega} (1-H(\phi_j)) (c_{inside}^j - |v_n|)^2 dx \nonumber \\
																																					&&	+ \lambda_2 \int_{\Omega} (\prod_{j=1}^N H(\phi_j)) (c_{outside} - |v_n|)^2  dx.
\end{eqnarray}
This term will make sure the segmentation describes the data well, where $\lambda_1$ and $\lambda_2$ are parameters to influence the importance of the two parts of the data term. However, simply minimizing $E_{data}$ would yield a segmentation with many unconnected components like a k-means clustering. Even single isolated pixels that have a high normal velocity due to noise would be assigned as a part of a cell. To get a more meaningful segmentation we need to take spatial information into account. We regularize with the length of the segmentation contour to obtain larger patches and use the idea that two adjacent pixels are likely to belong to the same classification group, foreground or background. The length of the contour can be measured by
\begin{eqnarray}
E_{reg}(\{\phi_j\}_{j=1,...N}) = \mu \sum_{j=1}^{N}  \int_{\Omega} \delta(\phi_j) |\nabla \phi_j| dx,
\end{eqnarray}
where $\delta(\phi)$ denotes the Dirac-Delta distribution. This term punishes long contours and its importance is determined by the parameter $\mu$. Thus, if the length of the contour is large in relation to the enclosed volume, $E_{reg}$ will dominate $E_{data}$ and small parts will vanish.

We introduce a third term that stabilizes the tracking algorithms and makes it more robust to frames for which the cell does not move. In this case the first two terms would simply force the contour to vanish since the normal velocity of the cell would be zero. However, we can include our a-priori information that the cell undergoes only little changes from one frame to the next and definitely does not vanish. Hence, we constrain the total cell volume,$\int_{\Omega} (1-H(\phi_j)) dx$, to stay close to the cell volume in the previous frame, $V^j_{old}$, by using the term
\begin{eqnarray}
E_{vol}(\{\phi_j\}_{j=1,...N}) = \nu \sum_{j=1}^{N}  \big( \int_{\Omega} (1-H(\phi_j)) dx- V^j_{old} \big)^2.
\end{eqnarray}

Our total energy then reads as follows
\begin{eqnarray}
E = E_{data}+E_{reg}+E_{vol}.
\end{eqnarray}

Our goal is to find the argument that minimizes the total energy. Therefore, we make use of the necessary condition for extrema that the first variation of the energy with respect to the quantity we are minimizing for has to be zero, which yields
\begin{eqnarray}
\frac{\delta E}{\delta c_{inside}^j} &=&2 \lambda_1 \int_{\Omega} (1-H(\phi_j)) (c_{inside}^j - |v_n|) dx = 0, \\
\frac{\delta E}{\delta c_{outside}}&=&2 \lambda_2 \int_{\Omega} (\prod_{j=1}^{N} H(\phi_j)) (c_{outside} - |v_n|) dx = 0, \\
\frac{\delta E}{\delta \phi_j} &=&  - \lambda_1 \delta(\phi_j) (c_{inside}^j - |v_n|)^2 \nonumber \\
												&&	+ \lambda_2 \delta(\phi_j) (c_{outside} - |v_n|)^2 \nonumber \\
												&& -  \mu \delta(\phi_j) \nabla \cdot \frac{\nabla \phi_j}{|\nabla \phi_j|} \nonumber \\
												&& - 2\nu \delta(\phi_j)\big( \int_{\Omega} (1-H(\phi_j)) dx- V^j_{old} \big)  = 0.
\end{eqnarray}

We solve the resulting equations  in an alternating way. The optimal $c_{inside}^j$ and optimal $c_{outside}$ for given $\phi_j$ can be calculated directly. For the minimization with respect to $\phi_j$ we use a gradient descent method. An artificial time variable is introduced and steps in the direction of the steepest descent, $-\frac{\delta E}{\delta  \phi_j}$, are taken until we approximate the minimum well. We iterate
\begin{eqnarray}
c_{inside}^j &=& \frac{\int_{\Omega} (1-H(\phi_j)) |v_n| dx}{\int_{\Omega} (1-H(\phi_j))dx}, \\
c_{outside} &=& \sum_{j=1}^{N} \frac{\int_{\Omega}  (\prod_{j=1}^{N} H(\phi_j))|v_n| dx}{\int_{\Omega}  (\prod_{j=1}^{N} H(\phi_j)) dx }, \\
\frac{\partial \phi_j }{\partial t} &=& - \frac{\delta E}{\delta \phi_j},
\label{phipde}
\end{eqnarray}
until (\ref{phipde}) reaches a steady state. Numerically, we use an explicit discretization of (\ref{phipde}) and smooth approximations for the delta and Heaviside functions, similar as in \cite{ChanVese} 

Since only the zero isocontour of the level set functions is important for the cell segmentation it is sufficient to update the $\phi_j$ only in a small neighborhood of their isocontours and omit all other values in the calculation. This trick of speeding up a level set algorithm is called narrow band method (\cite{Adalsteinsson95}) since all calculations are done on a narrow band around $\{ x | \phi(x) =0 \}$. The narrow band is updated as soon as the contour gets too close to the narrow bands border. This method is particularly well suited for our algorithm since it can easily be combined with the topology preservation we will discuss in Subsection \ref{topologyconstformultiplelevelsets}.

In our energy we have several free parameters to adapt to different type of data. We can weight the cost of excluding pixels from the segmentation with $\lambda_1$ and of including by $\lambda_2$. The smoothness of the contour is determined by $\mu$ and the importance of preserving the volume by $\nu$. In our experiments we used $\lambda_1 =4$, $\lambda_2 =2$, $\nu=0.03$, and $\mu = 3$. Choosing four parameters might seem difficult at first glance. However, the parameter only have to be tuned once for a certain type of cell video. The results of the five experiments with MDCK-F cells (result section, table \ref{comptable}), were all obtained by exactly the same set of parameters. Typically, experimentalists will focus on only a few types of different cells and microscopes, such that the tuning for a certain type of data set is not too laborious. Furthermore, the tuning of the parameters follows certain rules. For instance, $\mu$ regulates the smoothness of the curve and removes parts with little volume. Hence, the bigger the cells to be tracked are and the smoother/less deformed they are, the higher one should choose $\mu$. The parameter $\nu$ weights the volume preserving term. The penalty of this term scales with the cell volume and $\nu$ should therefore be chosen smaller the bigger the cell is. Finally, $\lambda_1$ and $\lambda_2$ determine how many pixels should be ex- and included. They should depend on the movement of the cells. If we have strong cell movement, the cells will clearly be visible in the normal velocity image and we would not prefer ex- or including over one another. As the movement gets smaller and there is more noise or missing movement, it makes sense to choose $\lambda_1 > \lambda_2$ to be sure to fully surround the cell. With these basic rules the parameter tuning for each video is a limited effort. Moreover, the manual segmentation on the first frame can be used to suggest a reasonable set of parameters, based on the volume and the smoothness of the contour of the manually segmented cells. 
 
\subsection{Topology Preservation}
\label{topologyconstformultiplelevelsets}
Despite the excellent results the above segmentation algorithm gives on the normal velocity there is still one major tracking problem we need to address. Level set methods provide a very flexible framework for image segmentation and allow any type of topology changes. This means that during the segmentation the contour of each level set function can split, merge, or move into regions of other cells. This problem is well known in cell tracking and is often referred to as undersegmentation (two adjacent cells wrongly get connected) and oversegmentation (a single low contrast cell gets split into several parts).

\begin{figure}[ht]
			\centering
			\includegraphics[width=15cm]{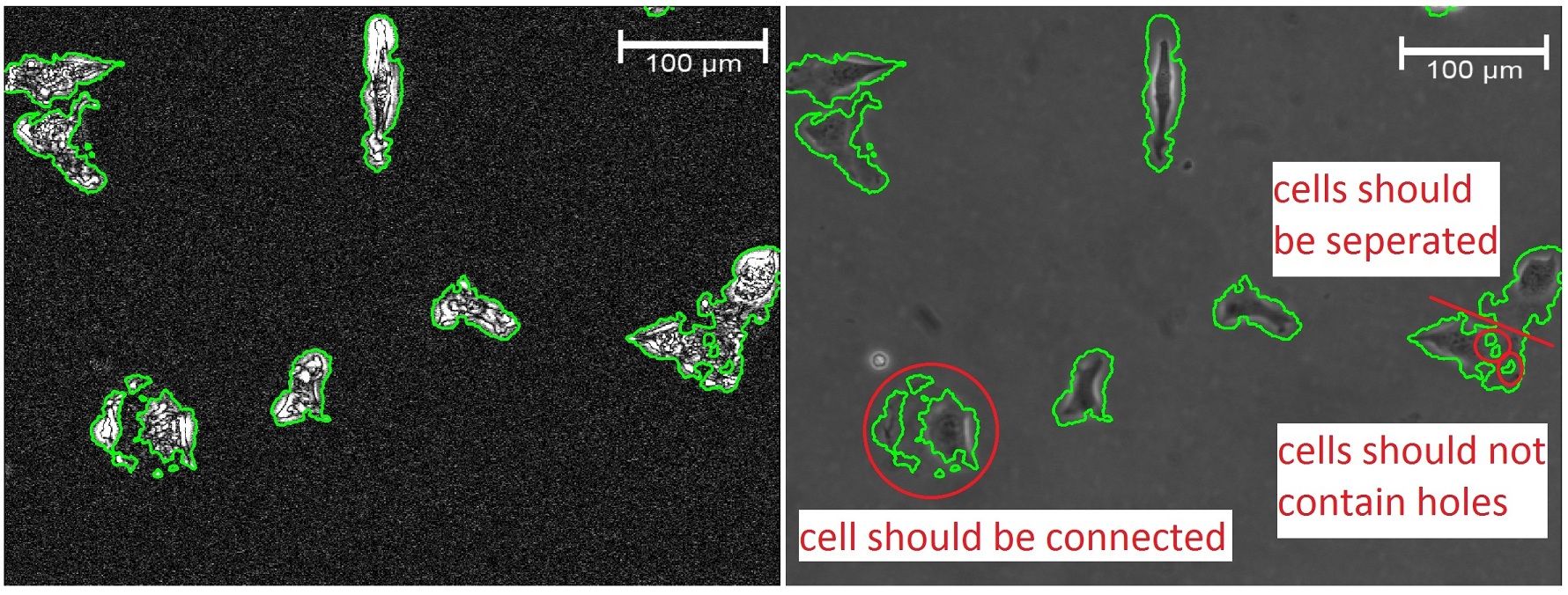}
			\caption{Over- and undersegmentation. Left: segmentation on the normal velocity field, right: segmentation and errors on corresponding video frame.}
			\label{wrongtopologysegementation}
\end{figure}

A typical example of these effects can be seen in Figure \ref{wrongtopologysegementation} which shows the segmentation result of the standard Chan-Vese method on a normal velocity image. We can see that several topological misclassifications appear, which could cause serious trouble for the tracking process. However, we know a-priori that for our data cells do not split, merge or create holes. It therefore makes sense to introduce this a-priori information and preserve the topology of each cell (no splitting or creating holes) and of the union of cells (no merging of two cells).

\begin{figure}[H]
			\centering
		\includegraphics[width=15cm]{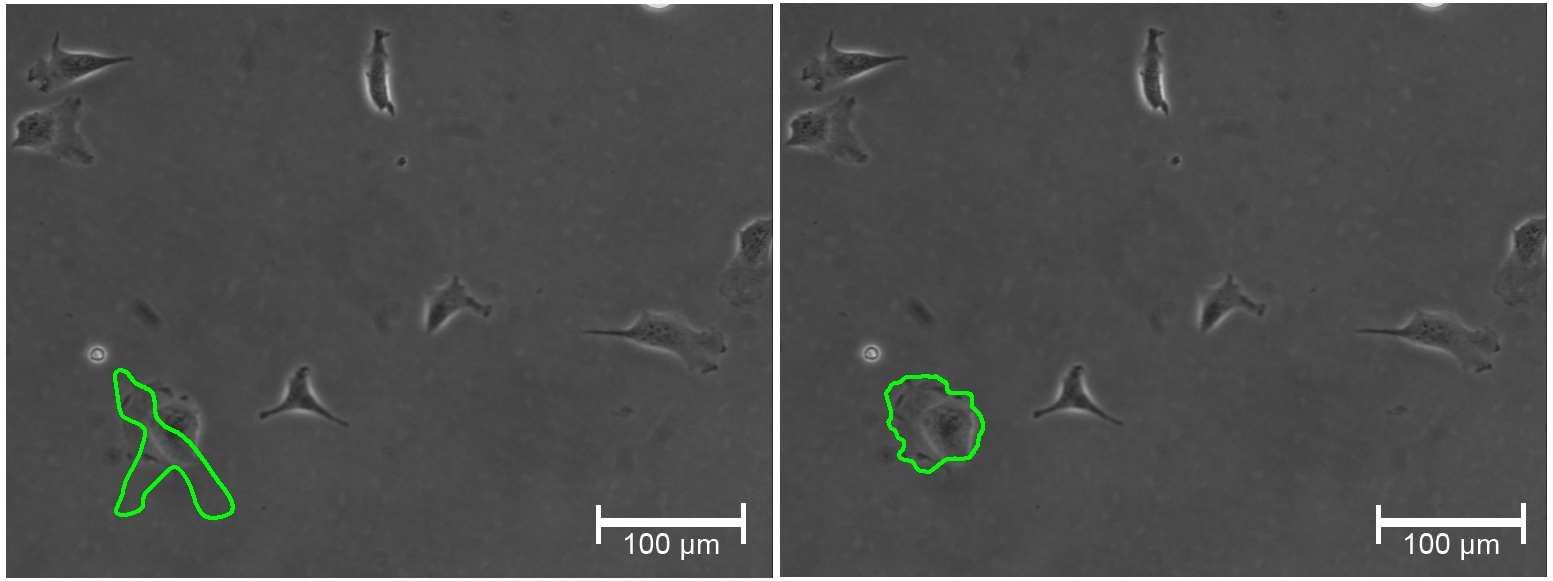}
			\caption{Dependence of the tracking on the initial contour. Left: initial contour, right: result of the normal velocity tracking}
			\label{initcontdependence}
\end{figure}

To preserve the topology during the tracking process we of course need one segmentation that definitely has the correct topology. Thus, we decided to implement a semi-automatic tracking algorithm in which the user has to mark the cells in the first frame of the video manually. This initial segmentation can be very rough as long as it captures the right topology. Each cell should have one contour without holes that does not intersect with the contours of the other cells. Figure \ref{initcontdependence} shows an example of an initial contour far off the actual cell boundary (left image). After running the normal velocity tracking algorithm we obtain the contour shown in the right image which is a good approximation of where the cell is. Thus, after refining the segmentation based on the normal velocity, i.e. after moving the contour inwards particularly at the upper left and middle right part of the cell, we will have captured the position of the cell very well - despite the bad initialization we started with. 

 Starting with the initial contours we segment each normal velocity image under the constraint of topology preservation meaning cells are not allowed to merge or split. This is done on the pixel level similar to the topology preserving geodesic active contours proposed in \cite{topologypreservation}. 

Notice that since we assign an own level set function for each cell we could work with a repulsion term as in \cite{ZhangMultiLS, NathMultiphaseLS, ZimmerSummary} to prevent cells from merging. However, the topology constraint also prevents each single cell segmentation from splitting or creating holes. Since we are interested in the shape of the cells, which we will measure by the ratio of cell area and the square of the perimeter, holes or split cells would lead to large shape index errors.

In \cite{topologypreservation} Han et al. proposed an algorithm to preserve the topology in geodesic active contour methods which are based on a (single) level set representation. Their idea was to use the gradient descent equation, in our case Equation (\ref{phipde}), as a temporary update first by calculating
\begin{eqnarray}
\phi^{temp} = \phi^{old} - \Delta t \frac{\delta E}{\delta \phi}.
\end{eqnarray}
The topology of the segmented object is only affected at pixels where the sign of the potentially new level set function, $\phi^{temp}$, is different from the former representation of the contour, $\phi^{old}$. If the sign has not changed we will update $\phi$ as usual, setting $\phi^{new} = \phi^{temp}$. If the sign of $\phi^{temp}$ is different from the sign of $\phi^{old}$ we will need to check if this change of sign changes the topology. It is shown in \cite{topologypreservation, detaileddigitaltopology} that there is a relatively easy criterion to check whether the digital topology of a contour has changed. This criterion only depends on the $3 \times 3$ neighborhood of a pixel and is compatible with a narrow band implementation for level set methods, which gives good algorithm speed. For details regarding digital topology preservation we refer to \cite{topologypreservation}.

\begin{figure}[ht]
			\centering
			\includegraphics[width=15cm]{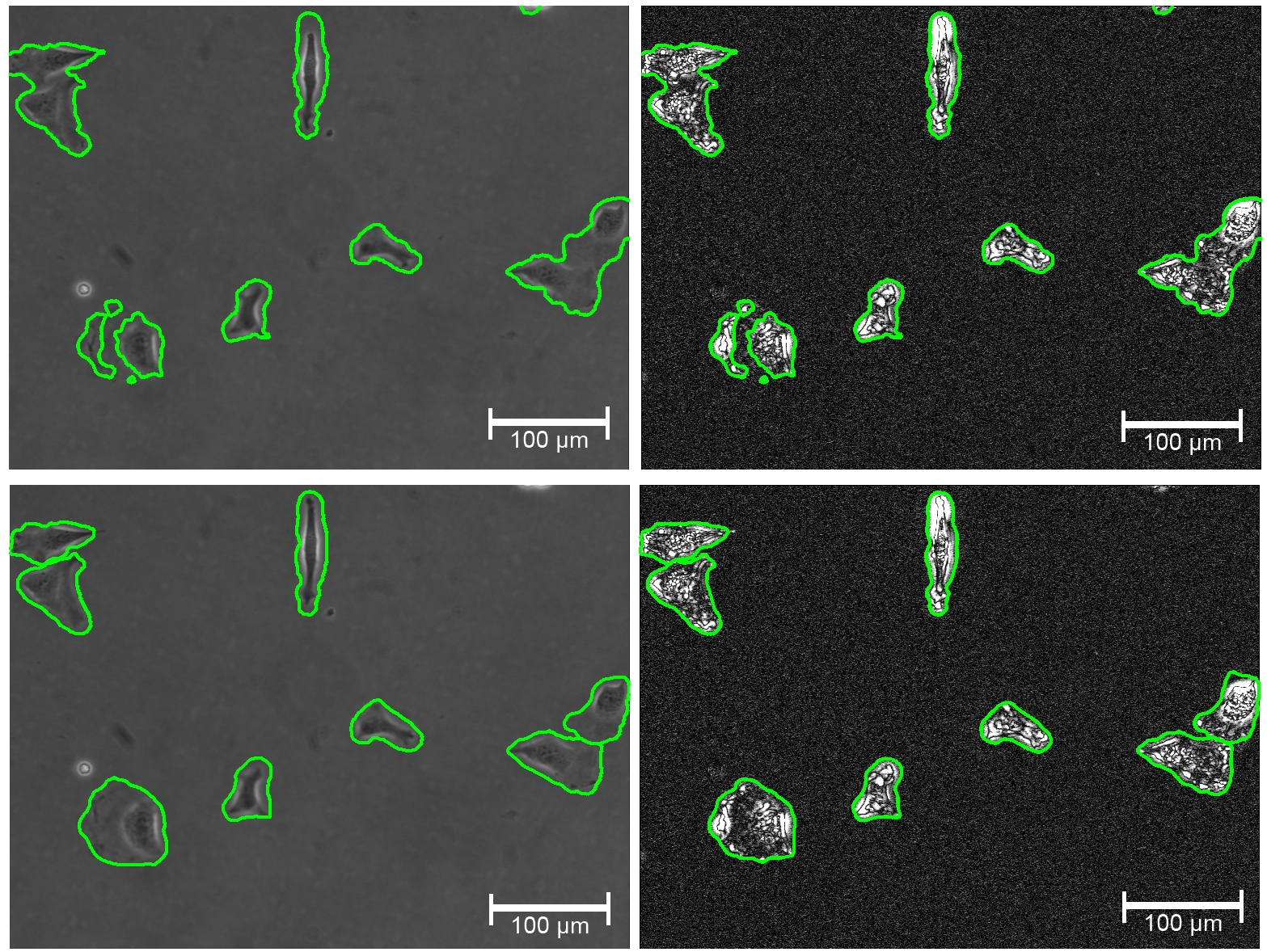}
			\caption{The effect of topology preservation. Top: Chan-Vese, bottom: topology preserving Chan-Vese with volume constraint}
			\label{topocomparison}
\end{figure}

Figure \ref{topocomparison} shows a comparison of the standard Chan-Vese segmentation result on a cell image and the result of our topology preserving algorithm with the volume constraint. We can see that cells that are very close to each other are connected in the standard model while the topology constraint in our algorithms prevents the contours from merging. Once the cells have been initialized by the user they will never change topology. For any kind of parameter choice the segmentation will not split or merge with other cells. We thereby avoid over- and undersegmentation as in Figure \ref{wrongtopologysegementation}. Furthermore, by having an own level set function for each cell, we avoid having to connect the single segmentations on the frames. A connection model is still implicitly made by initializing the contour in the next frame by the final contour from the previous one. The underlying assumption is that a cell does not move a lot between two successive frames. For our datasets this is a very reasonable assumption and a common methodology in general (see for instance \cite{Dzyubachyk08}). We will see the time resolution of our data set in the results section (figure \ref{successive}) which demonstrates that the contour from the previous frame can be used as a good initialization. 

\subsection{Contour Refinement}
The bottom row of Figure \ref{topocomparison} shows a typical example for the results of our first, rough tracking step. Since we mainly process the difference between successive frames we can only expect our segmentation to be as precise as a typical distance of cell movements between two frames. Due to little shifts in cell halos or variations of intensity around the cell borders the first segmentation step usually is even less precise and only an approximation of the true cell boundary. However, all previously described effects will lead to a too large segmentation. It therefore is a reasonable assumption to say that we have to move inwards until we hit the actual cell boundary to improve our segmentation. 

The remaining problem is to find a good criterion to detect the actual cell boundary. Locally cells seem to have a higher variance than the background. However, there are parts of the cell that have very low contrast to the background. Our idea is to create a new image based on the Laplacian of a Gaussian-smoothed cell image (similar to the edge detection proposed in \cite{marrhildreth}) and make the contour move inwards until it reaches areas of high Laplacian values. To prevent the contour from moving inwards too far at low contrast areas we define `high Laplacian values' for each cell individually and enforce a smooth contour in our model. Mathematically speaking we evolve the following geodesic active contours PDE (as it can be found e.g. in eq. (6) in \cite{topologypreservation})
\begin{eqnarray}
\frac{\partial \phi}{\partial t} = g(x) \kappa(x,t) | \nabla \phi(x,t) | + \nabla g(x) \cdot \nabla \phi(x,t).
\end{eqnarray}
We do the described procedure for each cell, i.e. for each $\phi_i$ separately. For the sake of simplified notation we leave out the index $i$ in the following. The first term moves the curve by its mean curvature $\kappa(x,t)= \nabla \cdot \frac{\nabla \phi}{| \nabla \phi |}$ and slows down or stops when $g$ gets close to zero. The second term passively moves the contour in the direction of larger decreases of $g$. For each cell, i.e. for each level set function from the previous segmentation and for each frame of the video, we design an individual $g$ by
\begin{enumerate}
	\item smoothing the current frame with a small Gaussian kernel,
	\item calculating the absolute value of the Laplacian of the smoothed image and
	\item calculating $\tilde{g}$ by two thresholds: we set $\tilde{g}=2$ for the lowest $70\%$ of the Laplacian values and $\tilde{g}=-2$ for the highest $26\%$ with a smooth transition for the intermediate values.
	\item To avoid the influence of the possible halo around the cell, we additionally set $\tilde{g}=2$ for too large gray values. 
	\item Finally, $g$ is calculated as the convolution of $\tilde{g}$ with a small Gaussian kernel.
\end{enumerate}

The threshold of $70\%$ and $26\%$ for the cacalculation of $g$ were determined experimentally. We calculate g in the following manner. Let $n$ be the number of pixel inside the $1-H(\phi_i)$.
\begin{eqnarray}
c &:=& \text{sort}\Big(|\Delta \big( (G \ast f)(1-H(\phi_i)) \big)|, \text{ ascending} \Big) \\
 n_1 &:=& 0.7 n, \ \  n_2 := 0.74 n \\
\tilde c_i &=& \left\{\begin{array}{cl} 2, & i<n_1  \\ -2 + 4 (i-n_1)/(0.04n), & n_1 \leq i \leq n_2 \\  -2 & i \geq n_2   \end{array}\right. \\
\tilde{g_i} &=& \left \{ \begin{array}{cl} c_i, & \mbox{ if $f_i<$threshold }    \\ 2, & \mbox{else} \end{array}\right. \\
g &=& G \ast \text{backsort}(\tilde{g})
\end{eqnarray}

\begin{figure}[H]
			\centering
			\includegraphics[width=15cm]{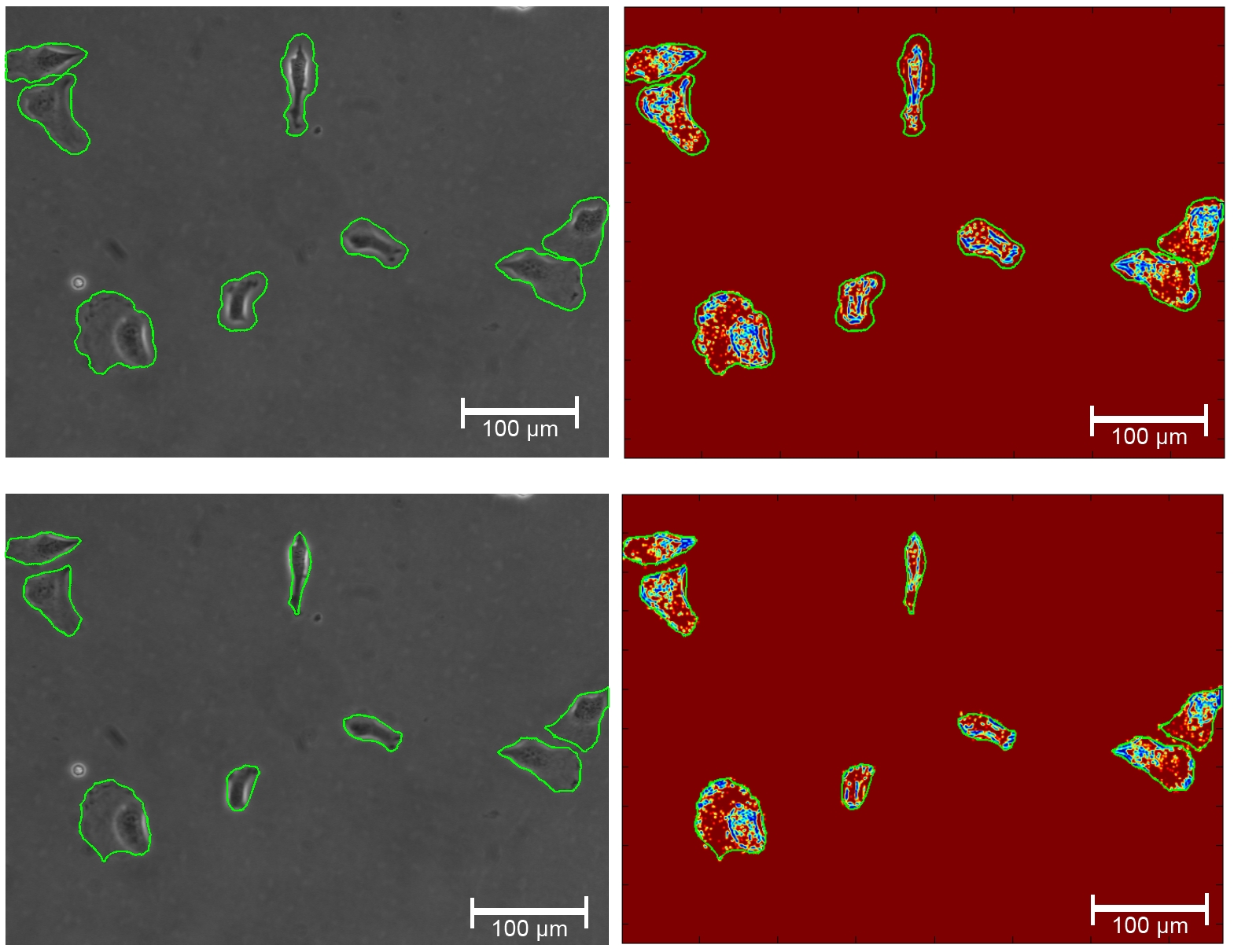}
			\caption{Left: contours before and after the refinement step. Right: corresponding $g$ whereas red represents large values and blue represents low values which the contour cannot pass.}
			\label{gmap}
\end{figure}

Figure \ref{gmap} shows an example of a cell and its corresponding $g$ before and after the contour refinement. We can see that the cell boundaries are reflected by the refined contour more accurately although it is hard to determine the true cell boundary on some parts. For some cells the halo has falsely been included. The method proposed in \cite{ErsoyNormalDirectionRefinement} explicitly addresses the problem of segmenting the halo and suggests to move the contour inwards until the gray value changes from bright to dark. However, in the experiments with our data this was too restrictive and gave too small cell boundaries. 

Let us summarize our method. Step 1: initialization of the cells by the user. Although automatic methods could be used instead, the first frame is the one that will provide the topology for the whole tracking which is why we prefer a manual initial contour. Notice that this initial contour does not have to be very accurate. Most important is the correct topology. Basically, we can start with a circle or an ellipsis around each cell. 
Step 2: Rough tracking based on the normal velocities using a topology preserving Chan-Vese method with a topology constraint. This is the key point and our key contribution of the algorithm. The Chan-Vese algorithm provides a segmentation tool not based on edges or gradients in the image but on regional information which makes it robust towards noise and enables us to segment the normal velocity images. The topological flexibility of level set methods must be restricted for cell tracking applications. For experiments that intend to measure the cell area and shape of the cell over time it is extremely important to have a segmentation algorithm that does not allow a cell to split, merge or form holes in the interior. Thus, we introduce topology preservation. Furthermore, we need robustness to a few frames of no or little movement of the cell which is provided by our volume constraint. Only the combination of these ingredients provide a reliable tracking method.
Step 3: For a more precise estimate of the shape and area we refine our segmentation locally based on smoothed image gradients to detect the cells boundary. 

\subsection{Cells entering and leaving the field of view}
One issue we have not addressed yet is how to handle cells entering or leaving the video. Although there are not many changes in the number of cells in the video data we have, we at least want to discuss how these problems can be handled. 

The moving of cells out of the screen is rather simple. When a cells leaves the screen the level set function shrinks and finally reduces to a single pixel. Although it does not entirely vanish (due to the topology preservation) we can simply threshold and eliminate any contour that is smaller than a certain minimal volume. 

A more difficult issue is to deal with cells that are newly entering the field of view. To detect such cells, we use the normal velocity image on our current frame and replace all current cells with the average background value (which is due to noise). Then, we smooth the image with an average filter and run the standard Chan-Vese segmentation model. Having this segmentation, we fill in all holes in closed segmentation curves and label the unconnected components. Any component whose area is above a certain threshold is called a newly detected cell. One could but does not have to restrict the search of new cells to the boundary of the image. Notice that this procedure can also be applied for the initialization and works well for sufficiently separated moving cells. However, for the reasons described in Section \ref{topologyconstformultiplelevelsets} we think it is more reliable to use the manual initialization. For detecting new cells the above issues of the Chan-Vese model are less problematic because 
\begin{enumerate}
	\item Falsely connecting two adjacent cells is less likely to happen since they would have to enter the screen adjacently and simultaneously.
	\item Having several unconnected components for one cell will likely lead to each of the parts being below the area threshold and would then postpone the detection of this cell until the segmentation found a connected component.
	\item Holes in the segmentation are avoided by construction. 
\end{enumerate}
We found the method suggested above to work reasonably well, particularly for cells with good movement entering the video separately. 

\section{Results}
\label{results}
\begin{figure}[H]
			\centering
		\includegraphics[width=16cm]{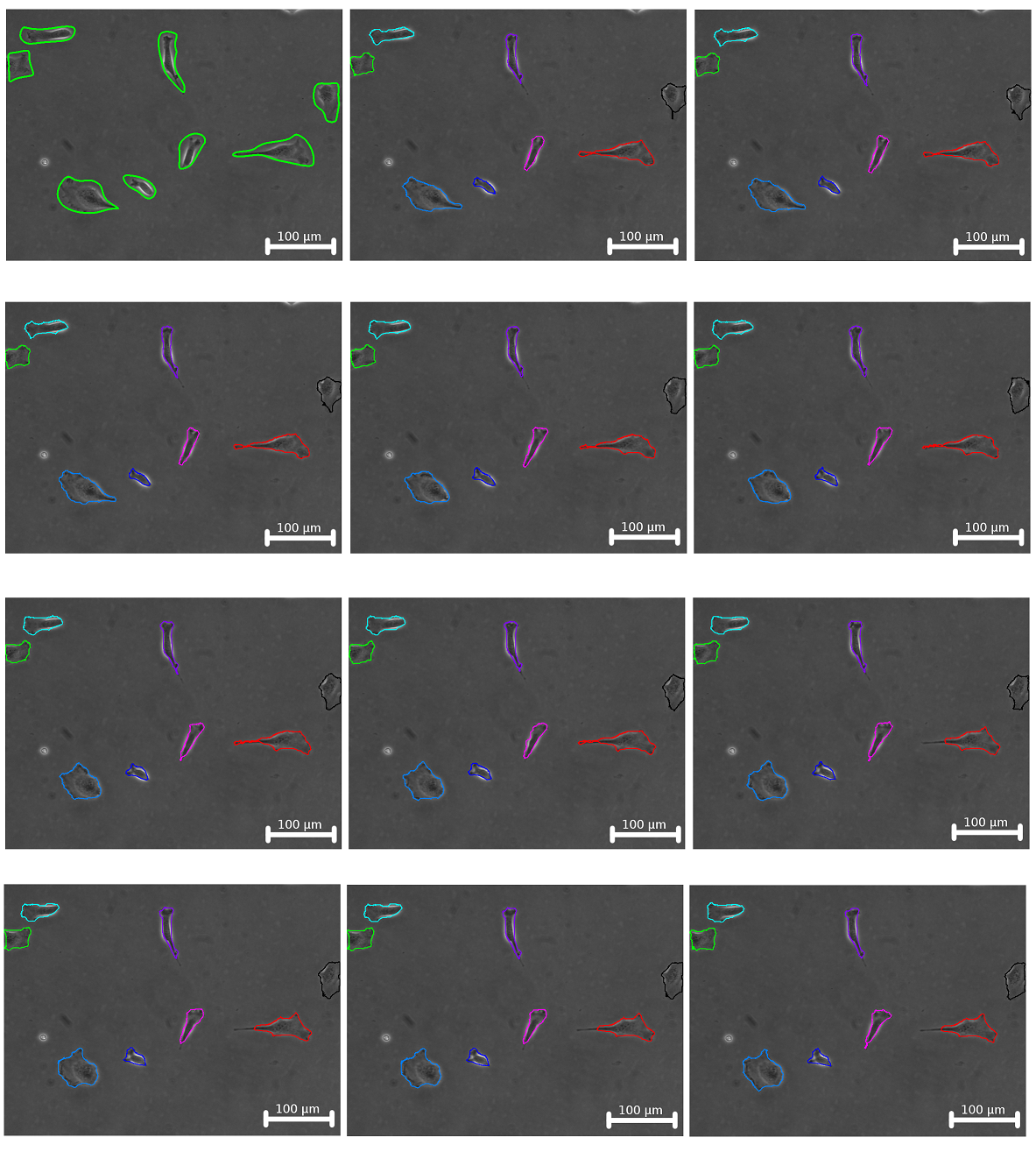}
			\caption{Successive frames of an example video with initialization (upper left) and tracking results of our algorithm.}
			\label{successive}
\end{figure}

First, let us show the tracking results on several successive frames to provide the possibility of visual inspection of our tracking results and to illustrate the time resolution of our video. Figure \ref{successive} shows nine successive frames ($\Delta t = 1$min) with the manual initialization (upper left) and the segmentation results of our algorithm. To underline that each cell has its own level set function we colored the contour of each cell differently. We can see that the time resolution is relatively high, which justifies the initialization of each new frame by the old contour. Our method describes most cell shapes well. Imprecisions occur due to halos, dirt particles or if the cells become very long and thin in which case the length term of our energy makes it difficult to stretch into long thin parts. However, overall we obtain good approximations of the true contours and more importantly do not lose cells or accumulate errors. 

\begin{figure}[H]
			\centering
		\includegraphics[width=15cm]{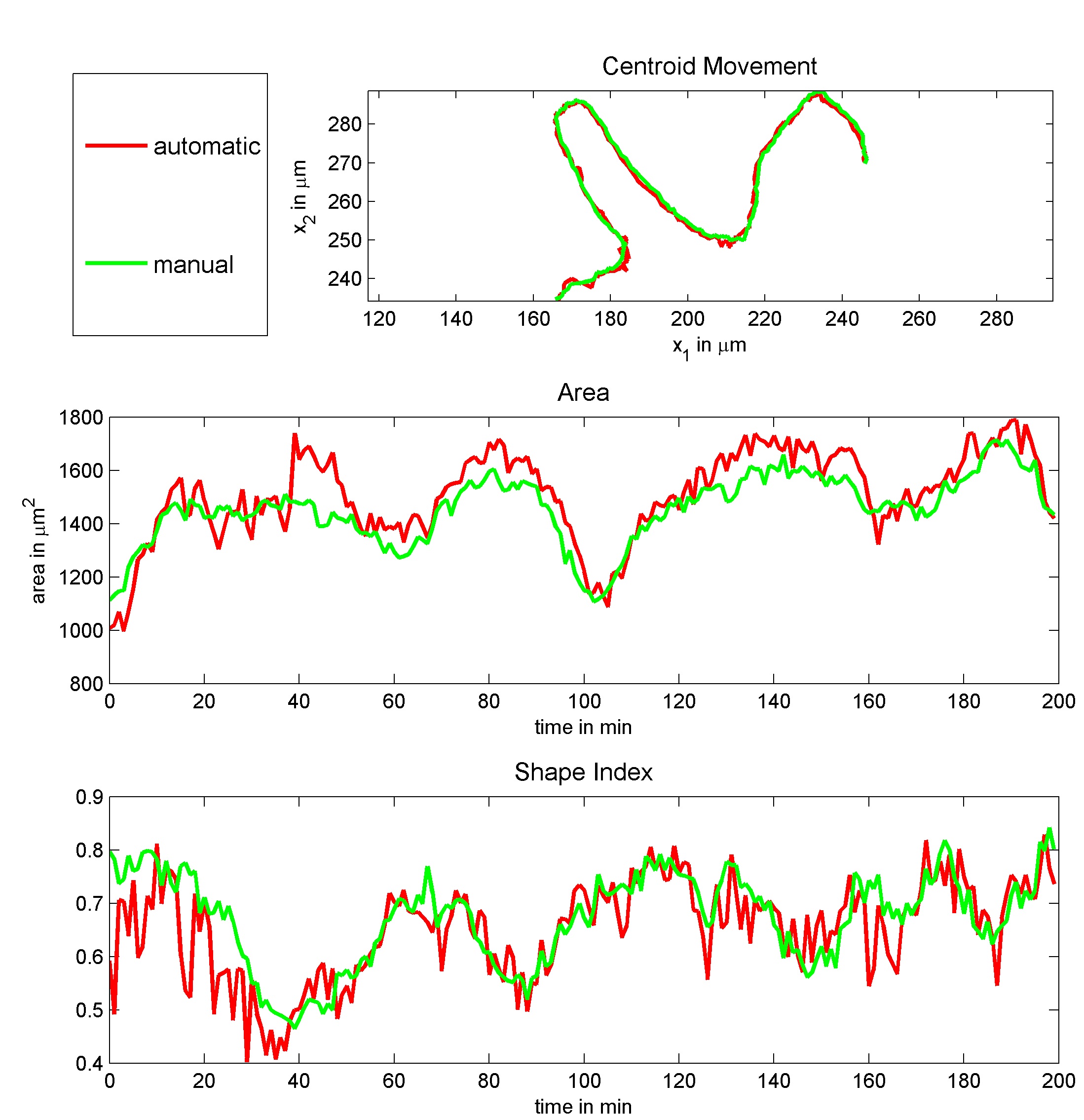}
			\caption{Comparison of manual and automatic segmentation}
			\label{firstcomparison}
\end{figure}

Visual inspection is of course not a sufficient evaluation of the algorithm. To evaluate our tracking method more accurately we made several experiments on MDCK-F cell data for which the tracking and corresponding analysis of the cell segmentation had been done manually. We determined the centroid position of the cells as well as their area, and a structural index defined as $SI = 4 \pi \frac{\text{Area}}{\text{Perimeter}^2}$ for each frame the cell was tracked in. Table \ref{comptable} shows the overall deviations from the manual segmentations for different cells in different experiments. For experiment 4 we had four different manual segmentations and compared our method to the average numbers of these four trackings.
\begin{table}
	\centering
		\begin{tabular}{|p{2cm} |c  | c  | c| c |c |c |c |c |}
		\hline
		 & & & time & & avg. er- & max. & avg. & \\
		 & & number & inter- & $\mu m$ & ror in & error in & area er- & avg. SI \\
		 experiment & cell & of & val in & per & centroid & centroid & ror in& error \\
		 name & & frames & min. & pix& pos. in & pos. in& $\%$ & in $\%$ \\
		 & & & & & pix. & pix. & & \\ \hline\hline
		
																			  	   &    1   &    125   &     1   &   0.675   &    2.19   &     7.31   &     4.91    &    10.71  \\ \cline{2-9}
			 																		   &   2   &   125   &   1   &   0.675   &   1.72   &   5.50   &   18.46   &   29.62  \\ \cline{2-9}
		 \raisebox{2.5ex}[-2.5ex]{experiment 1}  &   3   &   125   &   1   &   0.675   &   2.63   &   9.09   &   20.53   &   23.55  \\ \hline \hline
		
																			  	   &   1   &   200   &   1   &   0.675   &   1.79   &   5.28   &   5.23   &   7.16    \\ \cline{2-9}
			 																		   &   2   &   200   &   1   &   0.675   &   2.10   &   4.34   &   4.85   &   12.24  \\ \cline{2-9}
			 																		   &   3   &   200   &   1   &   0.675   &   2.27   &   8.74   &   9.41   &   10.96  \\ \cline{2-9}
		 \raisebox{3.5ex}[-3.5ex]{experiment 2}  &   4   &   200   &   1   &   0.675   &   1.65   &   4.60   &   5.70   &   14.51  \\ \hline \hline

			 																		   &   1   &   100   &   1   &   0.675   &   2.67   &   6.72   &   9.22   &   9.75  \\ \cline{2-9}
		 \raisebox{1.5ex}[-1.5ex]{experiment 3}  &   2   &   100   &   1   &   0.675   &   3.31   &   5.66   &   10.03   &   11.00  \\ \hline \hline

	 	\raisebox{0ex}[-0ex]{experiment 4}  &   1   &   31   &   1   &   0.65   &   2.13   &   5.76   &   3.07   &   13.29  \\ \hline \hline

			 																		    &   1   &   120   &   1   &   0.675   &   1.48   &   4.28   &   3.50   &   8.20  \\ \cline{2-9}
		 \raisebox{1.5ex}[-1.5ex]{experiment 5}  &   2   &   120   &   1   &   0.675   &   1.86   &   4.89   &   5.61   &   8.90  \\ \hline \hline
		
	 	\textbf{average}   &     &     &     &     &   \textbf{2.09}  &   \textbf{9.09}  &  \textbf{8.29}  &  \textbf{13.06}  \\ \hline
	
		\end{tabular}
			 \caption{Deviation between manual and automatic segmentation}
		 \label{comptable}
\end{table}
Shown are the number of cells and the number of frames the cells were tracked in as well as the following errors:
\begin{itemize}
	\item average centroid error which we calculated by $\frac{1}{n} \sum_i \sqrt{(x^i_{manual}-x^i_{automatic})^2+(y^i_{manual}-y^i_{automatic})^2}$, where $n$ denotes the number of frames,
	\item maximum centroid error calculated by $\text{max}_{i} \big \{ \sqrt{(x^i_{manual}-x^i_{automatic})^2+(y^i_{manual}-y^i_{automatic})^2}\big \}$,
	\item average area error in percent calculated by $\frac{100}{n} \sum_i \frac{| A^i_{manual} - A^i_{automatic}|}{A^i_{manual}}$,
	\item average shape index error in percent calculated by $\frac{100}{n} \sum_i \frac{| SI^i_{manual} - SI^i_{automatic}|}{SI^i_{manual}}$.
\end{itemize}
Notice that we gave all errors either in percent or in pixel to be independent of the scale of the microscopic images. The average centroid error is roughly in the range of 1 to 3 pixels, whereas the maximum error varies stronger having a maximum value of about 9 pixel. The overall average area error is 8.3$\%$ mainly due to two greatly overestimated cells in the first experiment. Similarly, the shape index is also off at these cells. Deviating by 13$\%$  the shape index seems to be the least reliable quantity of the automatic tracking. This is partly due to the sesitivity of the perimeter to local variations as well as the uncertainties due to the halos. However, the pure numbers do not really show whether our algorithm is useful for practical purposes. It makes sense to look at one experiment with significant cell movement and examine the manually and automatically determined statistics for all frames. Figure \ref{firstcomparison} shows the comparison between the manual and automatic tracking of cell 1 in experiment 2 with respect to centroid movement, area and structural index. 

We can see in all graphs that our tracking algorithm reflects the general behavior of the cell very well. Particularly, the centroid movement of the manual and automatic tracking are almost identical. The average deviation of about two pixels from the manually determined position is very small with respect to the total cell movement of about 144 pixel, such that the general behavior and the path the cell took can be seen equally well in both curves. The cell area is not as precise as the centroid movement. We can see that the automatically determined cell area fluctuates more. However, we do capture the strong increase in cell area at the beginning as well as the minimum around frame 100. Only frames 40-50 show a significant misestimate of the cell area. This could be due to dirt particles or clutter in the background close to the actual boundary of the cell causing the contour refinement to stop too early. The shape index of our tracking is clearly noisier than the manual one. Still we can see that the behavior of the two curves is the same.

From a numerical point of view it would make sense to smooth the data and reduce the effect of noise. However, this would lead to the question of the right amount of data regularity and furthermore come with the drawback that the statistical data would not correspond to a particular segmentation anymore. Finally, smoothing the data might obscure oscillatory behavior that is intrinsic to many migrating cells, e.g. \cite{Barnhartetal10} . Therefore, we decided to show the raw data in this results section. 

We have seen that we obtain a very robust tracking of the centroid position. The cell area as well as the shape index are more difficult to determine precisely. The errors we see can have multiple reasons like little cell movement (such that the first tracking step fails), halos, or low background to cell contrast as well as dirt particles in the background (such that the refinement step becomes imprecise). Nevertheless, the most important question is whether we are in an acceptable error range with our segmentation. 

\begin{figure}[H]
			\centering
		\includegraphics[width=15cm]{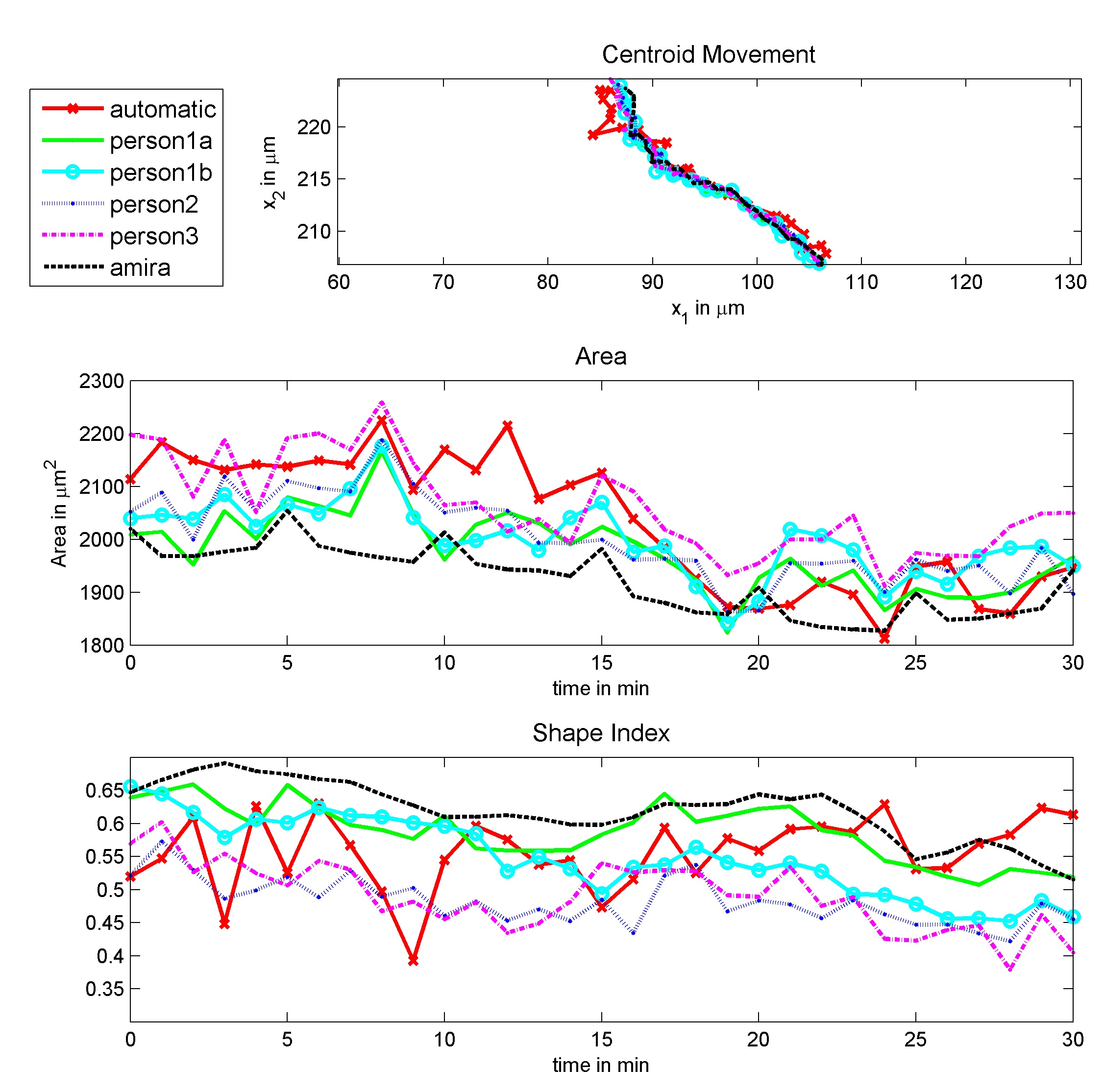}
			\caption{Comparison of manual and automatic segmentation}
			\label{secondcomparison}
\end{figure}

To investigate this question we were provided with test data for which three different persons independently tracked a cell over 31 frames. The first person did the tracking twice to see not only the user dependence but also the uncertainty one has for a single person's segmentation. Furthermore, an interpolation technique was applied where only every 5th frame was segmented manually and the frames in between  were interpolated by the AMIRA program, \cite{amira}. Figure \ref{secondcomparison} shows the comparison of all tracking methods. These images are taken from experiment 3 in the above Table \ref{comptable}. 

First of all we see the difficulty of precise cell tracking even for experienced investigators. Although the cell centroids of the manual tracking methods are relatively similar, the deviations in area and shape are rather high. This cell experiment seems to be more challenging for our automatic tracking algorithm. The centroid movement is not as well determined as in the previous example. However, the general behavior is still well reflected by our method. In the middle part of the tracking the automatically determined centroid position perfectly coincides with the manual ones. The main problem are the first couple of frames where even the manual methods show a larger variance than for the rest of the tracking. Looking closely at the video we can see a dirt particle close to the actual cell boundary that the algorithm classified as part of the cell.

The comparison of different manual segmentations shows that the area is hard to determine precisely. However, except for frames 10 to 14 the results of our algorithm seem to be in the range of uncertainty one typically has from the subjectiveness of the cell boundary. 

The behavior of the shape index is similar to the one of the cell area. We can also see relatively large differences between the tracking of different persons and even a clear difference between the two segmentations of the same person (blue and green line). Our algorithm yields a slightly higher shape index than most of the manual ones which is, however, in the typical error range of manual segmentations. Comparing the manual segmentations among themselves in terms of the errors we used for the evaluation above we get for instance when comparing person1a and person3 an average centroid error of 1.29 pixel, a maximum centroid error of 2.5 pixel, an average area error of 4.2$\%$ and an average shape index error of 20.4$\%$, which again indicates that the cell area and shape index seem to be in the range of uncertainty in manual trackings. 

Notice that we tested our method on real data which therefore contains a noise level typical for the imaging device. If the data contained substantially more noise while keeping the low contrast, preprocessing like for instance total variation denoising or non-local means would be necessary to reduce the noise level, since the calculation of the normal velocity is sensitive to strong noise. 

%

\begin{figure}[ht]
			\centering
			\includegraphics[width=15cm]{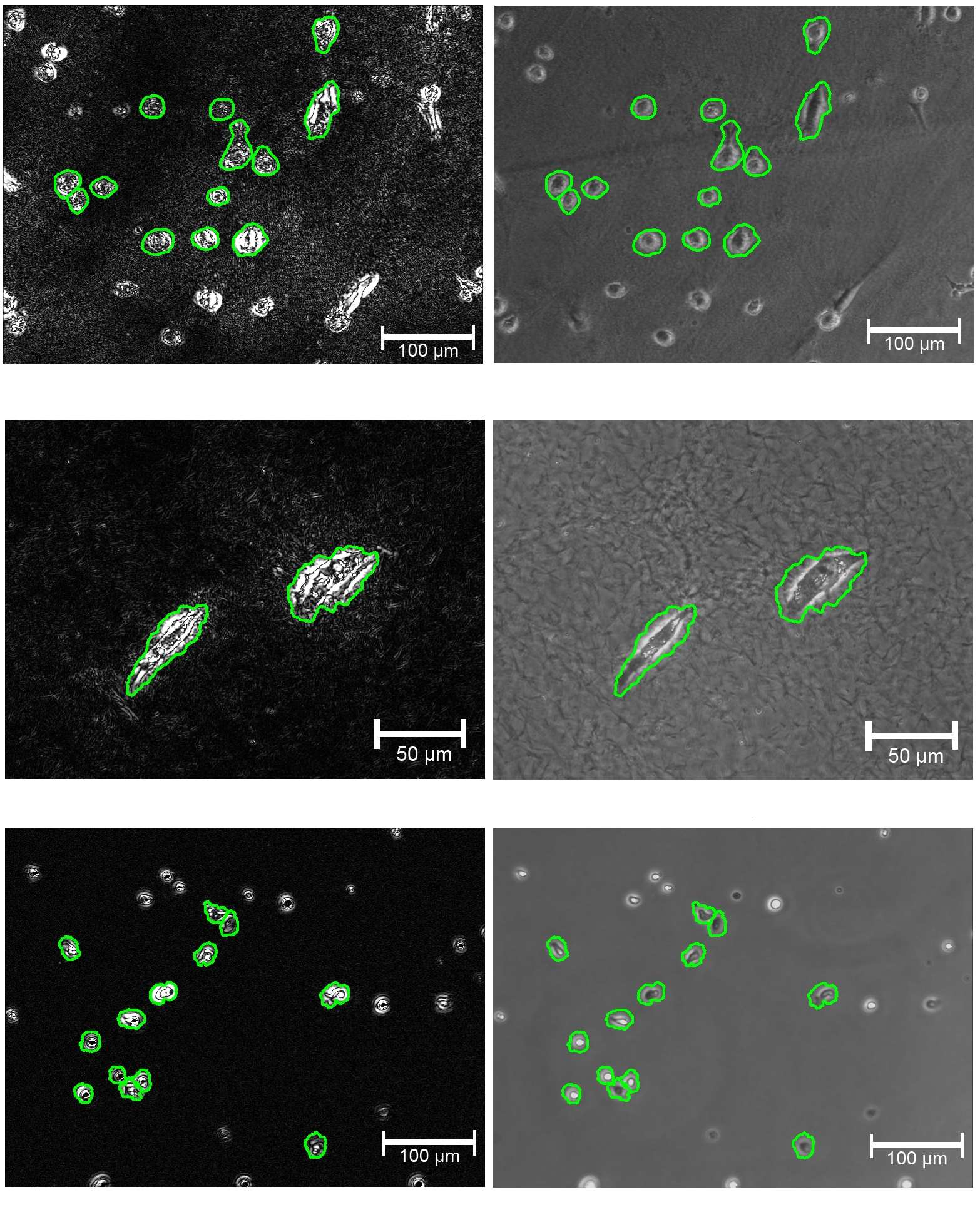}
			\caption{Human melanoma (MV3) cells migrating on a collagen I matrix (top and middle) and murine neutrophil granulocytes migrating on a fibronectin-coated culture dish. Segmentation results for the topology preserving Chan-Vese algorithm with volume constraint are shown.}
			\label{othercells}
\end{figure}

The general concept of tracking cells by their normal velocity with the help of a volume constrained topology preserving Chan-Vese algorithm can be extended to other types of cell data. Figure \ref{othercells} shows results of our algorithm on melanoma cells (top and middle) and murine neutrophil granulocytes (bottom). The cells in the normal velocity image are visible despite the substantial change of image types.

In the first image many small cells were tracked. Notice that the cell in the lower left image can barely be seen in the normal velocity image. This can happen if a cell stops moving for several frames. Thanks to the volume constraint the algorithm does not entirely loose this cell and will pick up the tracking once the cell starts moving again. 

The second type of image shows melanoma cells at a higher magnification on a nonuniform background. Due to the lower time resolution ($\Delta t = 10$min) the movement from frame to frame is more pronounced than for MDCK-F cells described in the previous sections.  Therefore, much larger bright spots appear in the normal velocity image leading to a rough segmentation. However, the centroid of the cells is still captured well by this kind of segmentation. The contour refinement needs to be adapted for these cells since the background is nonuniform and therefore the Laplacian is not a good criterion for the cell boundary any more. 

In the bottom image some cells appear dark and some appear white. Nevertheless the normal velocity is a feature that does not depend on the specific gray value and we obtain a good cell segmentation. For this data the time resolution was very high such that we use frames which were six time steps apart to calculate the time derivative. Again a contour refinement for this type of cell would have to be developed individually.

Besides the evaluation of accuracy we also want to point out the time advantages the user has with our algorithm. The complete automatic cell tracking as shown above takes less than one minute per cell per frame on a 2.5 GHz Intel Dual Core Processor with 4GB memory and therefore the time expenses are only in the order of magnitude of the manual tracking. The algorithms are all programmed in Matlab and are not particularly speed optimized. After drawing the initial contour the program does not need to be supervised and can therefore easily run overnight, such that the runtime is not a critical component. On the contrary, manual tracking is extremely time consuming for the observer. Furthermore, the automatic algorithm can handle arbitrarily many cells at the same time. In the manual segmentation software we have, one can only segment single cells. Therefore, the observer has to go through the whole stack of frames multiple times, if multiple cells are to be tracked. These significant advantages of the automatic software can save a lot of money and time and confirm the need of automatic cell tracking algorithms, particularly for large data volumes. 

\section{Conclusions}
\label{conclusion}
Our method seems to be able to provide a robust semi-automatic tracking method for phase contrast time lapse videomicroscopic migration experiments. On videos with a given manual tracking as a ground truth we stay in a reasonable range from the correct segmentation. Particularly, for long time experiments the centroid movement can be determined equally well with our proposed algorithm. The area and the structural index show larger variance as well as a larger deviation from the manual segmentations. However, we have seen that even the manual trackings show large differences in these quantities, which proves the subjectiveness of the cell boundaries and underlines the need for a method with reproducible results. To reduce the high variance of the automatically determined quantities show one could smooth the resulting data. On the other hand, this would come at the cost of not having an exact contour representing the data points. 


\section*{Acknowledgements}
AS acknowledges support form the grants IZKF M\"{u}nster Schw2/30/08 and DFG Schw407/9-3. MM and MB acknowledge support from the German Ministery for Science and Technology (BMBF) under project "Segmentation and Cartoon Reconstruction in Optical Nanoscopy". Furthermore, the authors would like to thank Christian Stock and Otto Lindemann for providing data sets of migration experiments with human melanoma cells and murine neutrophil granulocytes, respectively.

\bibliographystyle{alpha}
\bibliography{references}

\end{document}